\documentclass{aa}  

\usepackage{natbib}
\bibpunct{(}{)}{;}{a}{}{,} 
\usepackage{graphicx}
\usepackage{txfonts}
\usepackage[colorlinks=true,linkcolor=blue,citecolor=blue, urlcolor=blue]{hyperref}
\usepackage{xcolor}

\newcommand{\teff}{$T_{\rm eff}$}
\newcommand{\eexc}{$E_{\rm exc}$}
\def\vt{$\xi_{\rm t}$}
\def\kms{$\rm km~s^{-1}$}
\def\ione{\,{\sc i}}
\def\ii{\,{\sc ii}}

\newcommand{\eps}{\log\varepsilon}

\begin{document}

\title{
Distinct barium isotope ratios in CEMP-s and CEMP-rs stars
\thanks{
Based on data obtained from the European Southern Observatory Science Archive and the Keck Observatory Archive.}}
\titlerunning{Ba isotopes in CEMP-s and CEMP-rs stars}
\authorrunning{Sitnova et al.}

\author{T. M. Sitnova\inst{1}\thanks{sitamih@gmail.com}, L. I. Mashonkina\inst{1}, A. M. Romanovskaya\inst{1}, R. E. Giribaldi\inst{2}, A. Choplin\inst{3}}
\institute{
Institute of Astronomy, Russian Academy of Sciences, Pyatnitskaya 48, 119017, Moscow, Russia\
\and
INAF -- Osservatorio Astrofisico di Arcetri, Largo E. Fermi 5, 50125 Firenze, Italy\
\and
Institut d'Astronomie et d'Astrophysique, Universit\'e Libre de Bruxelles,  CP 226, B-1050 Brussels, Belgium
           }

\date{Received ; accepted }

\abstract
{}
{
We present a spectroscopic analysis of ten carbon enhanced metal-poor (CEMP) stars of type CEMP-s and CEMP-rs and determine their non-local thermodynamic equilibrium abundances of Ba and Eu, as well as the fractions of the odd Ba isotopes (F$_{\rm odd}$).
Determination of F$_{\rm odd}$ in stars provides unambiguous information about the s-process contribution to their total Ba abundance. 
We aim to obtain observational constraints on the enrichment scenarios of CEMP-rs stars.
}
{
The Ba abundances inferred from the resonance Ba\ii\ 4554 and 4934~\AA\ lines depend on the adopted Ba isotope mixture. We perform calculations for different F$_{\rm odd}$ from 0.1 to 1.0 and determine the corresponding abundances from the Ba\ii\ resonance lines in each sample star. In addition, we determine the Ba abundances from the Ba\ii\ subordinate lines, which are almost independent of F$_{\rm odd}$. We then compare the Ba abundances derived from the subordinate lines with those from the Ba\ii\ resonance lines.
}
{
We found different F$_{\rm odd}$ values in CEMP-s and CEMP-rs stars. 
CEMP-s stars exhibit F$_{\rm odd}$ = 0.05$_{-0.03}^{+0.07}$, 0.17$_{-0.14}^{+0.63}$, 0.19$_{-0.14}^{+0.50}$, and 0.19$_{-0.12}^{+0.33}$. The obtained values agree, within the error bars, with the s-process F$_{\rm odd}$ = 0.10 and the solar F$_{\rm odd}$ = 0.18. Although the uncertainties are large, in three of four stars, the possibility of Ba isotopes origin in a pure r-process with F$_{\rm odd}$ = 0.75 can be excluded. 
CEMP-rs stars show F$_{\rm odd}$ = 0.34$_{-0.21}^{+0.55}$, 0.36$_{-0.14}^{+0.23}$, 0.44$_{-0.22}^{+0.43}$, 0.53$_{-0.38}^{+0.47}$, and  0.57$_{-0.31}^{+0.43}$, which are higher compared to those in CEMP-s stars. Although the uncertainties are large, in four of five stars,  the possibility of a pure s-process origin for the Ba isotopes can be excluded. The obtained values agree, within the error bars, with the predicted i-process F$_{\rm odd}$ = 0.6 to 0.8.
}
{
Our analysis of CEMP-rs stars with [Ba/Eu] > 0  argues that their [Ba/Eu] and F$_{\rm odd}$ cannot be jointly explained by a mixture of material produced by the r- and s-processes. The obtained results argue that the i-process is responsible for the chemical composition of these CEMP-rs stars.
}


\maketitle

\section{Introduction}

Carbon enhanced metal-poor (CEMP) stars  have been the subject of ongoing interest for several decades. Various mechanisms leading to carbon overabundances result in a vast subclassification of CEMP-type stars \citep{2005ARA&A..43..531B}. In this study, we focus on CEMP-s and CEMP-rs stars, which are binary system members \citep{2001AJ....122.1545P,2005ApJ...625..825L} and exhibit overabundances of neutron (n-) capture elements. These stars allow us to study the properties and nucleosynthesis of their intermediate-mass asymptotic giant branch (AGB) companions, which have now become white dwarfs.

Ba and Eu traditionally serve as key diagnostic elements and are known in the literature as slow (s-) and rapid (r-) process elements, respectively. In solar material, 88 \%\ of Ba originates from the s-process, and 95 \%\ of Eu from the r-process \citep{2020MNRAS.491.1832P}. Using the above values and the solar abundances of \citet{Lodders2009}, one can calculate [Ba/Eu]\footnote{We use a standard designation, [X/Y] = $\log($N$_{\rm X}$/N$_{\rm Y}$)$_{*} - \log($N$_{\rm X}$/N$_{\rm Y}$)$_{\odot}$, where N$_{\rm X}$ and N$_{\rm Y}$ are total number densities of elements X and Y, respectively.} = 1.25 and $-0.9$ for the pure s- and r-processes, respectively. In the literature, the [Ba/Eu] ratio is used to distinguish between CEMP-s and CEMP-rs stars, namely, CEMP stars with higher [Ba/Eu] are considered CEMP-s stars, and those with lower [Ba/Eu] are considered CEMP-rs stars. The exact [Ba/Eu] abundance ratio, along with other elemental ratios such as [Ba/Fe], [Eu/Fe], and others used for classification, varies between studies \citep{2005ARA&A..43..531B,2006A&A...451..651J,2016A&A...587A..50A,2018ARNPS..68..237F,2019A&A...623A.128H,2021A&A...645A..61K,2021A&A...649A..49G}. Numerous classification schemes available in the literature demonstrate that distinguishing between CEMP-rs and CEMP-s stars is not a trivial task.

Chemical element abundances in CEMP-s stars provide observational evidence for the enrichment with elements produced by the main s-process in AGB companion stars, while the nature of the n-capture element overabundances in CEMP-rs is a matter of debate. Various scenarios  have been proposed to explain the Eu overabundance in CEMP-rs stars along with the Ba overabundance. The first one implies that an r-process enhanced star has been contaminated with s-process material from its companion star \citep[see, for example,][]{2000A&A...353..557H,2006A&A...451..651J,2011MNRAS.418..284B}. The second one suggests a contamination of a star with material produced in the intermediate (i-) n-capture process \citep[]{2016ApJ...831..171H}. Its astrophysical site and physical conditions are not clearly understood. For example, i-process nucleosynthesis scenarios in MP asymptotic giant branch stars \citep{2022A&A...667A.155C,2024A&A...684A.206C} or white dwarfs \citep{2019MNRAS.488.4258D}  have been proposed. And, of course, these scenarios are not mutually exclusive, and all of them can exist and be responsible for the chemical composition of different CEMP-rs stars. For more details on the formation mechanisms of CEMP-rs stars, see the discussion in \citet{2021A&A...649A..49G}. 

To obtain observational constraints on the formation scenarios of CEMP-rs stars, accurate data on the abundances and isotope ratios in CEMP-rs stars are needed. Determination of the Ba isotope ratio in stars provides unambiguous information about the s-process contribution to the Ba abundance. In stars, Ba can be represented by five isotopes: $^{\rm 134}$Ba, $^{\rm 135}$Ba, $^{\rm 136}$Ba, $^{\rm 137}$Ba, and $^{\rm 138}$Ba. The isotopes $^{\rm 134}$Ba and $^{\rm 136}$Ba are s-only isotopes and cannot be produced in the r-process and i-process due to the presence of stable r-only isotopes $^{\rm 134}$Xe and $^{\rm 136}$Xe, which block the $\beta$-decay pathway and prevent the formation of $^{\rm 134}$Ba and $^{\rm 136}$Ba. For different n-capture processes, calculations predict different fractions of odd isotopes (F$_{\rm odd}$ = (N($^{135}$Ba)+N($^{137}$Ba))/N(Ba)). For example, F$_{\rm odd}$ = 0.10 \citep[s-process,][]{2020MNRAS.491.1832P}, 0.60 to 0.80 \citep[i-process,][]{2025EPJA...61...68C}, and 0.75 \citep[r-process,][]{2020MNRAS.491.1832P}. 
Therefore, the ratio of odd to even isotopes allows us to distinguish stars in which Ba is mainly produced by the s-process from other stars.

The method of Ba isotope ratio determination relies on the fact that the odd isotopes are subject to hyperfine splitting (HFS) of the energy levels, and a higher F$_{\rm odd}$ results in a broader line profile and a greater total absorbed energy. HFS primarily affects the ground state, meaning that the Ba\ii\ resonance lines can serve as a diagnostic of the Ba isotope ratio. In contrast, subordinate lines are unaffected by the adopted isotope ratio and can be used as reliable indicators of barium abundance. This feature can be used for F$_{\rm odd}$ determination by comparing abundances from the subordinate lines and the resonance lines computed with different F$_{\rm odd}$. This idea was first proposed by \citet{1989ApJ...346.1030C}, who pointed out the importance of taking HFS into account when determining Ba abundances. The abundance comparison method of F$_{\rm odd}$ determination was applied to MP stars in the Milky Way \citep{2006A&A...456..313M,2008A&A...478..529M,2019AstL...45..341M,2025A&A...699A.262S} and the Sculptor dwarf spheroidal galaxy  \citep[][]{2015A&A...583A..67J}. In this study, we aim to apply this method to CEMP-s and CEMP-rs stars.

To better understand the nature of CEMP-s and CEMP-rs stars, we are wondering:
(i) Do CEMP-s and CEMP-rs stars have distinct Ba isotope ratios?
(ii) Do Ba isotope ratios help to distinguish different enrichment scenarios of the CEMP-rs stars: i-process or a mixture of the r- and s- processes?
Our key diagnostics for identifying n-capture element sources are [Ba/Eu] and Ba isotope ratios.

The paper is structured as follows. In Sect.~\ref{sample_obs}, we describe our sample stars and observations. Stellar atmosphere parameters are presented in Sect.~\ref{stpar}. The abundance determination method is presented in Sect.~\ref{abund}. A discussion of our findings is provided in Sect.~\ref{discussion}, and our conclusions are summarised in Sect.~\ref{conclusions}.

\section{Stellar sample and observations}\label{sample_obs}

For our analysis, we selected stars known in the literature as CEMP-s or CEMP-rs stars. To perform Ba isotope ratio determinations, we set the following criteria: (i) the resonance lines and the subordinate lines of Ba\ii\ must be detectable in high-resolution spectra publicly available in the ESO and Keck archives; (ii) the Ba\ii\ resonance lines must be strong enough to trace Ba isotope ratios, while remaining not too saturated to allow for accurate abundance measurements; (iii) the Ba\ii\ lines must be free from blending with the carbon lines.

Selecting CEMP stars suitable for Ba isotope ratio analysis is challenging due to both high carbon and barium abundances in these stars. To minimise the impact of carbon line blending and Ba\ii\ lines saturation, we selected stars with high \teff\ and log~g -- dwarfs and subgiants (Table~\ref{sample}). In \citet{2025A&A...699A.262S}, we took stars with the Ba\ii\ resonance line equivalent width (EW) of less than 140~m\AA. Applying the same threshold to CEMP stars results in a very small sample of five stars, which does not allow to draw a reasonable conclusion. To double our sample, in this study we raised the upper limit to 180~m\AA. 
This threshold still allows us to stay within a reasonable saturation range in the growth curve and not to  deal with the upper atmospheric level effects \citep{2025A&A...702A..65G}. We analyse the sensitivity of the Ba\ii\ resonance lines to the isotope ratio as a function of EWs in Sect.~\ref{abund}.

We selected ten stars, which are listed in Table~\ref{sample} alongside characteristics of their observed spectra, stellar atmosphere parameters, and  classification from the literature. Five stars are known in the literature as CEMP-rs stars, two stars -- CEMP-s stars, and for the remaining three, it is unclear whether they are CEMP-s or CEMP-rs stars.

We use high-resolution and high signal-to-noise (S/N) ratio spectra available in the European Southern Observatory Science Archive and the Keck Observatory Archive. For six sample stars, we use data from the UV-Visual Echelle Spectrograph (UVES) at the UT2 Kueyen Telescope, we also use the High Accuracy Radial Velocity Planet Searcher (HARPS) spectrograph at the La~Silla~3.6m telescope, and the High-Resolution Echelle Spectrometer (HIRES) at the Keck~\ione\ Telescope. The adopted spectra are of reasonable quality, with a spectral resolution of $\lambda/\Delta \lambda >$ 30~000 and a signal to noise ratio per pixel of S/N $>$ 30. The program IDs are listed in Table~\ref{sample}. 

\begin{table*}
\caption{Stellar sample, atmospheric parameters, and characteristics of the observed spectra.} 
\setlength{\tabcolsep}{1.0mm}            
\label{sample}      
\centering          
\begin{tabular}{l l l l l l l l l l}   
\hline      
Name & Classification  &   \teff ,  & log g$^2$, & \tiny{[Fe/H]}        & \vt , &  \tiny{S/T$^3$} & \tiny{S/N$_{\rm red}$} & R,& Program ID \\
      &  ref.$^1$  & \tiny{K} & \tiny{$\rm cm~s^{-2}$} &     & \tiny{\kms} &       &    & 10$^3$&         \\
\hline                                                           
2MASS~J09124370+0216236  & rs (AP15)           &  6130 & 4.51 (0.04)           & --2.41 & 1.1 & U &  42 & 31 & \tiny{078.D-0217(A)               }  \\    
BPS CS 31062-012         & rs (AP15)           &  6200 & 4.44 (0.02)           & --2.60 & 1.2 & U & 114 & 65 & \tiny{105.20LJ.002                }  \\    
BPS CS 22881-036         & rs (AP15)           &  6140 & 4.38 (0.03)           & --1.87 & 1.2 & U & 114 & 52, 42 & \tiny{165.N-0276(A), 078.B-0238(A)}  \\
BPS BS 16080-175         & rs (AP15)           &  6210 & 4.11 (0.03)           & --1.46 & 1.4 & K & 146 & 48 & \tiny{U013Hr                      }  \\    
BD --01 2582             & rs (MS25)           &  5220 & 2.82 (0.03)           & --2.16 & 1.4 & U & 266 & 57 & \tiny{68.D-0546(A)                }  \\    
HD 196944                & s (AP15), rs (KV21) &  5390 & 2.13 (0.04)           & --2.21 & 2.0 & H & 224 & 115 & \tiny{60.A-9036(A)                }  \\   
2MASS J11140709+1828320  & s, rs (SC13)        &  6160 & 4.53 (0.08)           & --2.84 & 1.1 & U &  68 & 32 & \tiny{087.D-0123(A)               }  \\    
2MASS~J11432342+2020582& s, rs (SC13)        &  6390 & 4.53 (0.18)           & --3.06 & 1.2 & U &  58 & 32 & \tiny{087.D-0123(A)               }  \\    
HE 1029-0546             & s (HH15)            &  6440 & 4.05 (0.09)           & --3.22 & 1.5 & U &  46 & 57 & \tiny{077.D-0035(A)               }  \\    
HE 0450-4902             & s (HH15)            &  6570 & 4.44 (0.05)           & --2.56 & 1.4 & U &  32 & 35 & \tiny{077.D-0035(A)               }  \\    
\hline      
\multicolumn{10}{l}{1 -- Classification from the literature: AP15 -- \citet{2015A&A...581A..22A}, HH15 -- \citet{2015ApJ...807..173H}, MS25 --  }\\ 
\multicolumn{10}{l}{ \citet{2025arXiv251006968M}, KV21 -- \citet{2021A&A...645A..61K}, SC13 -- \citet{2013A&A...552A.107S}.}\\
\multicolumn{10}{l}{2 -- The log~g uncertainties are indicated in parenthesis. For each sample star, the uncertainties in \teff, [Fe/H], and \vt\ of 80~K,   }\\ 
\multicolumn{10}{l}{0.2~dex, and 0.1~\kms, respectively, were adopted.}\\
\multicolumn{10}{l}{3 -- Spectrograph/Telescope: U -- UVES/UT2, K -- HIRES/Keck\ione, H -- HARPS/La~Silla~3.6m.}\\ 
\end{tabular}
\end{table*}

\section{Stellar atmosphere parameters}\label{stpar}

We calculated effective temperatures (\teff ) using the {\it Gaia} ${\rm BP-G}$, ${\rm G-RP}$, ${\rm BP-RP}$ dereddened colours and the calibration of \citet{2021A&A...653A..90M}. The extinction ${\rm E(B-V)}$ is adopted from \citet{2011ApJ...737..103S} and the colours are corrected according to \citet{2018MNRAS.479L.102C}. Using different colours yields very similar effective temperatures, and the uncertainty in \teff\ is therefore mainly defined by an uncertainty in the calibration  of 80~K, as given by \citet{2021A&A...653A..90M}. This means that in the range of stellar parameters we deal with in this study, the molecular carbon bands at 4310~\AA\ are not strong enough to affect the colours.

We calculated surface gravities (log~g) using distances based on {\it Gaia} parallaxes. Parallaxes are corrected for the zero offset according to \citet{2021A&A...649A...4L} and distances are computed from the maximum of the probability distribution function as described in \citet{2015PASP..127..994B}. With these distances, effective temperatures, bolometric corrections of \citet{2018MNRAS.479L.102C}, and assuming a mass of 0.8 solar masses, we derive surface gravities using the formula $\log g = 4.44+\log(m/m_{\odot})+0.4(M_{\rm bol} - 4.75) + 4\log($\teff$/5780)$, where $m_{\odot}$ is a solar mass and $M_{\rm bol}$ is an absolute bolometric magnitude. 

Microturbulent velocity was computed using an empirical relation based on non-local thermodynamic equilibrium (NLTE) analysis of Fe\ione\ and Fe\ii\ lines in dwarfs \citep{2015ApJ...808..148S} and Fe\ione, Fe\ii, and Ti~\ii\ lines in giants \citep{2025A&A...699A.262S}. We estimated metallicity using LTE abundances from the strongest lines of Fe\ii\  4923~\AA\ and 5018~\AA. Since our sample stars are mostly hot dwarfs, many of them have no other Fe\ii\ lines except for these two. For the Fe\ii\ at 4923~\AA\ and 5018~\AA\ lines, we adopt the oscillator strengths log~gf =  $-1.39$ and $-1.23$, respectively, from \citet{1998A&A...340..300R} that were corrected by +0.11 dex, following the recommendation of \citet{1999A&A...347..348G}. The NLTE effects for the Fe\ii\ lines are negligible in the stellar parameter range considered in this study \citep{mash_fe,2019A&A...631A..43M}.

\section{Abundance analysis}\label{abund}
\subsection{Codes and model atmospheres}
We use classical 1D model atmospheres from the \textsc{marcs} model grid \citep{marcs}, interpolated for the given \teff, log~g, and [Fe/H] of the stars.
In this study, we deal with carbon-enhanced stars that are products of binary system mass-transfer. To make sure if using a standard model atmospheres for CEMP stars is reasonable, for one sample star, HD~196944, we computed two model atmospheres with different abundances of chemical elements. These calculations are performed with the \textsc{LLmodels} code \citep{2004A&A...428..993S}, which self-consistently accounts for individual elemental abundances in the opacity calculation. For one model, we use a standard chemical composition, while another one accounts for high carbon abundance [C/Fe] = 1.2 and modified  hydrogen and helium abundances adopted as N(H)/N$_{\rm tot}$ = 0.919 and $\eps$(He) = $-1.10$, respectively, in accordance with predictions for 2~m$_{\rm \odot}$ AGB star \citep{2015ApJS..219...40C} and the dilution factor of 0.8 \citep{2022A&A...662C...3C}. We found spectral lines under investigation to be similar in synthetic spectra computed with the two model atmospheres. A minor difference is detected in the Balmer line profiles due to a distinct hydrogen abundance.

We solve the coupled radiative transfer and statistical equilibrium equations using the \textsc{detail} code \citep{Giddings81,Butler84}, incorporating the updated opacity package presented by \citet{mash_fe}. For synthetic spectra calculations, we use the \textsc{synthV\_NLTE} code \citep{Tsymbal2018}, attached to the \textsc{idl binmag} code \citep{2018ascl.soft05015K}. This method allows us to obtain the best fit to the observed line profiles while accounting for the NLTE effects via pre-calculated departure coefficients (the ratio between NLTE and LTE atomic level populations) for a given model atmosphere. When fitting the line profiles, the abundance of the element of interest is varied alongside the macroturbulent velocity and the radial velocity.

The spectral synthesis line list is extracted from a recent version of the Vienna Atomic Line Database \citep[VALD,][]{2019ARep...63.1010P,2015PhyS...90e4005R}, which provides isotopic and hyperfine structure components for a number of studied chemical elements. The oscillator strengths for Ba\ii\ are taken from \citet{Miles_Wiese}, and for Eu\ii\ from \citet{LWHS}.

\subsection{Ba abundances}
We analyse five lines of Ba\ii: the strong resonance lines at $\lambda$ = 4554~\AA\ and 4934~\AA, and the subordinate lines at $\lambda$ = 5853~\AA, 6141~\AA, and 6496~\AA. In the vicinity of the Ba\ii\ 4934.0~\AA\ line and  6141.71~\AA\ lines, there are 
iron lines, namely, Fe\ione\ 4934.0~\AA\ with the lower level energy of \eexc\ = 4.15~eV and the oscillator strength of log~gf = $-0.58$ and Fe\ione\ 6141.73~\AA\ with \eexc\ = 3.60~eV and log~gf = $-1.46$. In both cases, blending with the Fe\ione\ lines in our sample stars is negligible, since they have too high \teff s to produce a significant absorption from these Fe\ione\ lines.

\begin{figure}
\centering
\includegraphics[trim={0 5 0 0},width=\hsize]{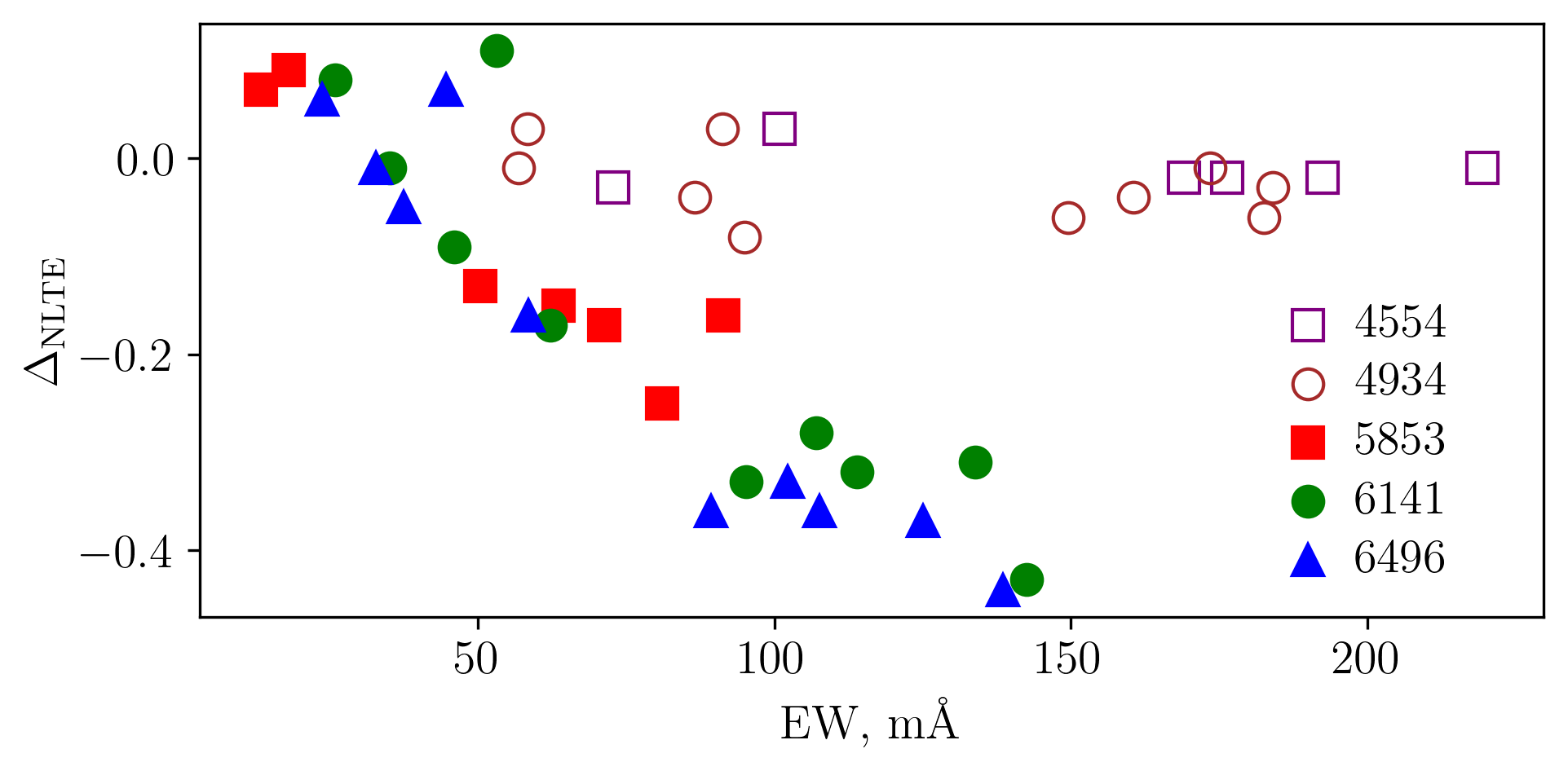}
\caption{NLTE abundance corrections for the Ba\ii\ lines as a function of EWs. See legend for lines designation.}
\label{ba_corr}
\end{figure}

To account for departures from LTE, we adopt the Ba\ii\ model atom from \citet{2019AstL...45..341M}. For the resonance lines, the NLTE effects are moderate, leading to either weaker or stronger lines compared to LTE, depending on the line strength and stellar atmosphere parameters. In our sample stars, the NLTE abundance corrections for the resonance lines do not exceed 0.1~dex in absolute value (Fig.~\ref{ba_corr}). For the subordinate lines, the NLTE abundance corrections are mostly negative and increase in absolute value, reaching $-0.45$~dex with the increasing EWs to 150~m\AA. 

In our spectral synthesis calculations, we consider five Ba isotopes: $^{\rm 134}$Ba, $^{\rm 135}$Ba, $^{\rm 136}$Ba, $^{\rm 137}$Ba, and $^{\rm 138}$Ba. 
We determine Ba abundances using different F$_{\rm odd}$: 0.10 (pure s-process), 0.18 (solar), 0.43, 0.62, 0.75 (pure r-process), and 1. Here,  Ba isotope yields in the s-process and the residual r-process are taken from \citet[][]{2020MNRAS.491.1832P}. For each sample star, we provide abundances from individual spectral lines together with their measured EWs (Table~\ref{ba_lbl}).

\begin{figure}
\centering
\includegraphics[trim={0 5 0 0},width=\hsize]{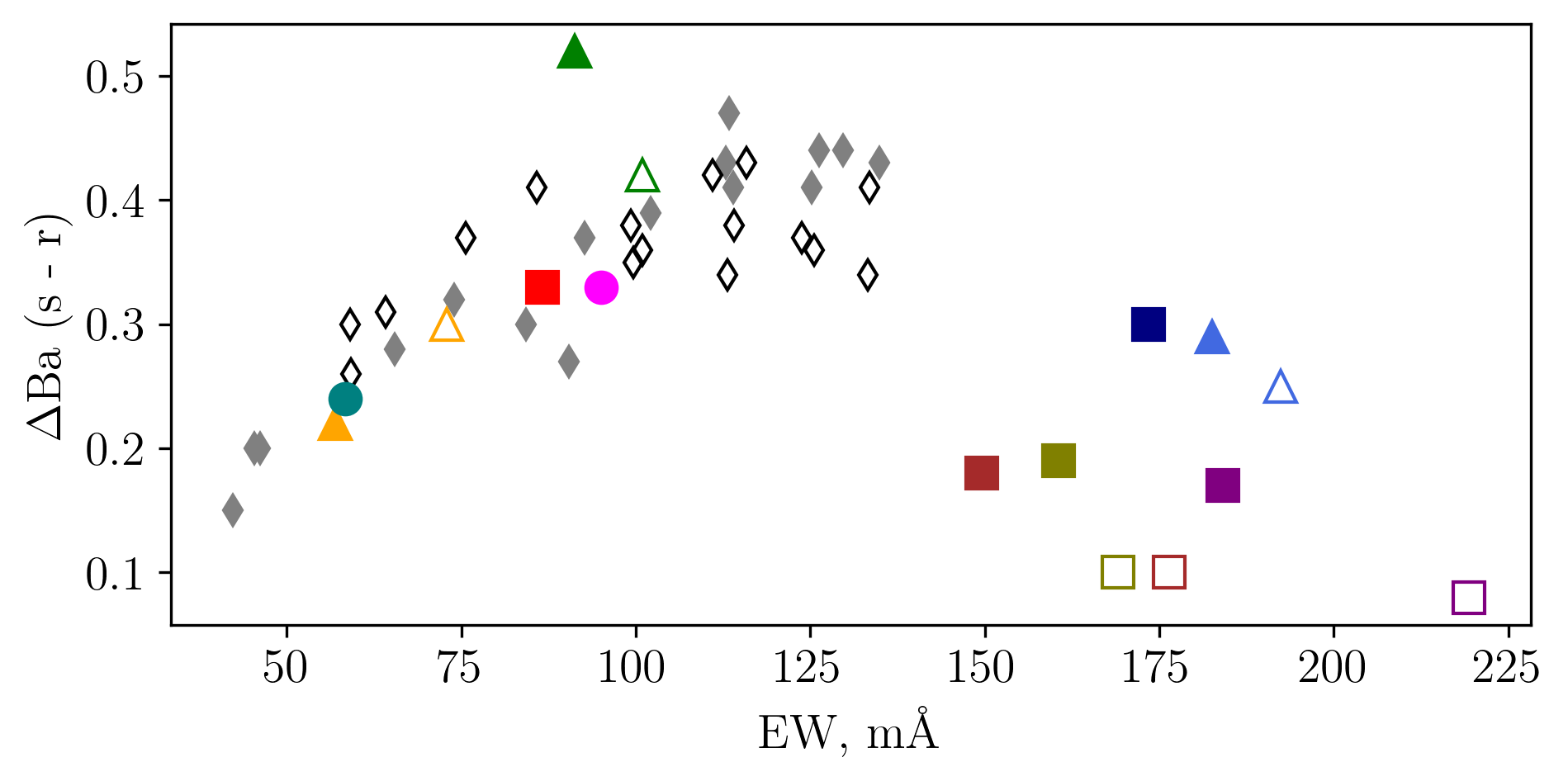}
\caption{
Differences between NLTE abundances from individual lines computed with F$_{\rm odd}$ = 0.10 (s-process) and 0.75 (r-process) in the sample stars known in the literature as CEMP-s (circles), CEMP-rs (squares), and with unclear classification (triangles). For comparison, we show data for normal VMP stars from  \citet[][diamonds]{2025A&A...699A.262S}. Open and filled symbols correspond to the Ba\ii\ 4554 \AA\ and 4934 \AA\  lines, respectively.
}
\label{ba_res_rs}
\end{figure}

Figure~\ref{ba_res_rs} illustrates the sensitivity of the Ba\ii\ resonance lines with different EWs to the adopted F$_{\rm odd}$. The impact of F$_{\rm odd}$ on Ba abundance increases with increasing EW, and the maximum difference in the NLTE abundances computed with F$_{\rm odd}$ = 0.10 (s-process) and 0.75 (r-process) of $\Delta$Ba(r-s) = 0.4 -- 0.5~dex is observed in stars with the Ba\ii\ EWs of 100 -- 140 m\AA. Further increase of EWs results in the line saturation, and stronger lines are less affected by the adopted F$_{\rm odd}$. The saturation effect becomes more pronounced for the Ba\ii\ 4554~\AA\ line, which has larger gf-value compared to the Ba\ii\ 4934~\AA\ line. For example, in BPS~BS~16080-175 the $\Delta$Ba(r-s) amounts to 0.08 and 0.17~dex for the Ba\ii\ 4554 \AA\ with Ew = 219~m\AA\ and 4934~\AA\ with EW = 184~m\AA, respectively. The former value is too small for F$_{\rm odd}$ determination, and, moreover, saturated lines provide less reliable abundance, regardless of the adopted F$_{\rm odd}$. The sensitivity of the resonance lines depends on the EW and also on stellar paremeters, therefore, it could be difficult to select a specific maximal EW. For our analysis, we selected the lines that exhibit a significant difference of more than 0.15~dex in abundances computed with the s-process and the r-process F$_{\rm odd}$. This abundance sensitivity threshold nearly corresponds to EW of 180~m\AA\ (Fig.~\ref{ba_res_rs}).

\begin{table*}
\caption{NLTE and LTE abundances and EWs (m\AA) of Ba\ii\ lines in the sample stars.}
\label{ba_lbl}
\centering   
\setlength{\tabcolsep}{1.0mm}      
   \begin{tabular}{lrrrrrrrrrrrrrrrr}    
      \hline
 Name & Ba\ii:    & \multicolumn{6}{c}{4554} & \multicolumn{6}{c}{4934} &  5853 &  6141  & 6496  \\
 &   & \multicolumn{6}{c}{-----------------------------------------------} & \multicolumn{6}{c}{-----------------------------------------------} &    &    &   \\
& \small{F$_{\rm odd}$:} & 1.0 & 0.75 & 0.62 & 0.43 & 0.18 & 0.10 & 1.0 & 0.75 & 0.62 & 0.43 & 0.18 & 0.10 & & &  \\
   \hline 
\tiny{J11432342+2020582 } & \tiny{NLTE} &    0.94 &    1.04 &    1.09 &    1.19 &    1.37 &    1.46 &    0.82 &    0.90 &    0.95 &    1.07 &    1.30 &    1.42 &    1.17 &    1.21 &    1.08 \\ 
\tiny{J11432342+2020582 } & \tiny{LTE } &    0.91 &    1.01 &    1.06 &    1.16 &    1.34 &    1.43 &    0.79 &    0.87 &    0.92 &    1.04 &    1.27 &    1.39 &    1.08 &    1.10 &    1.01 \\ 
\tiny{J11432342+2020582 } & \tiny{EW  } &   100.9 &   100.9 &   100.9 &   100.9 &   100.9 &   100.9 &    91.2 &    91.2 &    91.2 &    91.2 &    91.2 &    91.2 &    18.2 &    53.2 &    44.7 \\ 
\tiny{J11140709+1828320  } & \tiny{NLTE} &    0.23 &    0.26 &    0.29 &    0.36 &    0.50 &    0.56 &    0.18 &    0.18 &    0.19 &    0.24 &    0.34 &    0.40 &  --     &    0.53 &    0.63 \\ 
\tiny{J11140709+1828320  } & \tiny{LTE } &    0.26 &    0.29 &    0.32 &    0.39 &    0.53 &    0.59 &    0.19 &    0.19 &    0.20 &    0.25 &    0.35 &    0.41 &  --     &    0.54 &    0.64 \\ 
\tiny{J11140709+1828320  } & \tiny{EW  } &    72.9 &    72.9 &    72.9 &    72.9 &    72.9 &    72.9 &    56.9 &    56.9 &    56.9 &    56.9 &    56.9 &    56.9 &   --    &    35.2 &    32.9 \\ 
\tiny{J09124370+0216236 } & \tiny{NLTE} &  --     &  --     &  --     &  --     &  --     &  --     &    0.58 &    0.59 &    0.62 &    0.68 &    0.84 &    0.92 &    0.82 &    0.69 &    0.68 \\ 
\tiny{J09124370+0216236 } & \tiny{LTE } &  --     &  --     &  --     &  --     &  --     &  --     &    0.62 &    0.63 &    0.66 &    0.72 &    0.88 &    0.96 &    0.75 &    0.78 &    0.73 \\ 
\tiny{J09124370+0216236 } & \tiny{EW  } &   --    &   --    &   --    &   --    &   --    &   --    &    86.6 &    86.6 &    86.6 &    86.6 &    86.6 &    86.6 &    13.5 &    46.0 &    37.5 \\ 
\tiny{BPS CS 31062--012         } & \tiny{NLTE} &  --     &  --     &  --     &  --     &  --     &  --     &    1.68 &    1.70 &    1.73 &    1.77 &    1.85 &    1.89 &    1.61 &    1.51 &    1.52 \\ 
\tiny{BPS CS 31062--012         } & \tiny{LTE } &  --     &  --     &  --     &  --     &  --     &  --     &    1.72 &    1.74 &    1.77 &    1.81 &    1.89 &    1.93 &    1.74 &    1.84 &    1.88 \\ 
\tiny{BPS CS 31062--012         } & \tiny{EW  } &   169.1 &   169.1 &   169.1 &   169.1 &   169.1 &   169.1 &   160.5 &   160.5 &   160.5 &   160.5 &   160.5 &   160.5 &    50.4 &    95.3 &    89.4 \\ 
\tiny{BPS CS 22881--036         } & \tiny{NLTE} &    1.72 &    1.73 &    1.75 &    1.77 &    1.81 &    1.83 &    1.61 &    1.62 &    1.64 &    1.68 &    1.76 &    1.80 &    1.88 &    1.69 &    1.70 \\ 
\tiny{BPS CS 22881--036         } & \tiny{LTE } &    1.74 &    1.75 &    1.77 &    1.79 &    1.83 &    1.85 &    1.67 &    1.68 &    1.70 &    1.74 &    1.82 &    1.86 &    2.03 &    1.97 &    2.03 \\ 
\tiny{BPS CS 22881--036         } & \tiny{EW  } &   176.4 &   176.4 &   176.4 &   176.4 &   176.4 &   176.4 &   149.5 &   149.5 &   149.5 &   149.5 &   149.5 &   149.5 &    63.7 &   107.1 &   102.3 \\ 
\tiny{BPS BS 16080--175         } & \tiny{NLTE} &  --     &  --     &  --     &  --     &  --     &  --     &    2.02 &    2.04 &    2.05 &    2.09 &    2.17 &    2.21 &    2.13 &    2.05 &    2.03 \\ 
\tiny{BPS BS 16080--175         } & \tiny{LTE } &  --     &  --     &  --     &  --     &  --     &  --     &    2.05 &    2.07 &    2.08 &    2.12 &    2.20 &    2.24 &    2.38 &    2.36 &    2.40 \\ 
\tiny{BPS BS 16080--175         } & \tiny{EW  } &   219.4 &   219.4 &   219.4 &   219.4 &   219.4 &   219.4 &   184.1 &   184.1 &   184.1 &   184.1 &   184.1 &   184.1 &    81.1 &   133.9 &   125.1 \\ 
\tiny{HD 196944                } & \tiny{NLTE} &    0.88 &    0.93 &    0.96 &    1.02 &    1.13 &    1.18 &    0.66 &    0.74 &    0.76 &    0.80 &    0.95 &    1.05 &    0.87 &    0.77 &    0.78 \\ 
\tiny{HD 196944                } & \tiny{LTE } &    0.90 &    0.95 &    0.98 &    1.04 &    1.15 &    1.20 &    0.70 &    0.78 &    0.80 &    0.84 &    0.99 &    1.09 &    1.03 &    1.20 &    1.22 \\ 
\tiny{HD 196944                } & \tiny{EW  } &   192.4 &   192.4 &   192.4 &   192.4 &   192.4 &   192.4 &   182.6 &   182.6 &   182.6 &   182.6 &   182.6 &   182.6 &    91.4 &   142.5 &   138.6 \\ 
\tiny{HD 196944 test, G23      } & \tiny{NLTE} &    1.09 &    1.14 &    1.17 &    1.23 &    1.31 &    1.35 &    0.93 &    0.96 &    1.00 &    1.05 &    1.20 &    1.28 &    1.06 &    0.95 &    0.97 \\ 
\tiny{HD 196944 test, G23      } & \tiny{LTE } &    1.11 &    1.16 &    1.19 &    1.25 &    1.33 &    1.37 &    1.00 &    1.03 &    1.07 &    1.12 &    1.27 &    1.35 &    1.25 &    1.42 &    1.47 \\ 
\tiny{HD 196944 test, G23      } & \tiny{EW  } &   192.4 &   192.4 &   192.4 &   192.4 &   192.4 &   192.4 &   182.6 &   182.6 &   182.6 &   182.6 &   182.6 &   182.6 &    91.4 &   142.5 &   138.6 \\ 
\tiny{HE 1029--0546            } & \tiny{NLTE} &  --     &  --     &  --     &  --     &  --     &  --     &    0.17 &    0.21 &    0.24 &    0.29 &    0.40 &    0.45 &  --     &    0.35 &    0.42 \\ 
\tiny{HE 1029--0546            } & \tiny{LTE } &  --     &  --     &  --     &  --     &  --     &  --     &    0.14 &    0.18 &    0.21 &    0.26 &    0.37 &    0.42 &  --     &    0.27 &    0.36 \\ 
\tiny{HE 1029--0546            } & \tiny{EW  } &   --    &   --    &   --    &   --    &   --    &   --    &    58.4 &    58.4 &    58.4 &    58.4 &    58.4 &    58.4 &   --    &    25.9 &    23.8 \\ 
\tiny{HE 0450--4902            } & \tiny{NLTE} &  --     &  --     &  --     &  --     &  --     &  --     &    0.92 &    0.95 &    0.98 &    1.05 &    1.20 &    1.28 &  --     &    1.15 &    1.22 \\ 
\tiny{HE 0450--4902            } & \tiny{LTE } &  --     &  --     &  --     &  --     &  --     &  --     &    1.00 &    1.03 &    1.06 &    1.13 &    1.28 &    1.36 &  --     &    1.32 &    1.38 \\ 
\tiny{HE 0450--4902            } & \tiny{EW  } &   --    &   --    &   --    &   --    &   --    &   --    &    95.0 &    95.0 &    95.0 &    95.0 &    95.0 &    95.0 &   --    &    62.3 &    58.6 \\ 
\tiny{BD --01 2582             } & \tiny{NLTE} &  --     &  --     &  --     &  --     &  --     &  --     &    0.72 &    0.76 &    0.80 &    0.85 &    1.00 &    1.06 &    0.89 &    0.77 &    0.76 \\ 
\tiny{BD --01 2582             } & \tiny{LTE } &  --     &  --     &  --     &  --     &  --     &  --     &    0.73 &    0.77 &    0.81 &    0.86 &    1.01 &    1.07 &    1.06 &    1.09 &    1.12 \\ 
\tiny{BD --01 2582             } & \tiny{EW  } &   --    &   --    &   --    &   --    &   --    &   --    &   173.4 &   173.4 &   173.4 &   173.4 &   173.4 &   173.4 &    71.3 &   114.0 &   107.6 \\ 
\hline
      \end{tabular}\\
\end{table*}

\subsection{Eu abundances}

Our Eu abundance determination mainly relies on the Eu\ii\ 4129~\AA\ and 4205~\AA\ lines. For Eu abundance determination in BD~$-$01~2582 and 2MASS~J09124370+0216236, we had to use the Eu\ii\ 3819~\AA\ line, as the Eu\ii\ 4129~\AA\ and 4205~\AA\ lines are not covered with the available spectrum of BD~$-$01~2582 or too weak to be measured in the spectrum of 2MASS~J09124370+0216236. The Eu\ii\ 3819~\AA\ line is difficult to analyse, since it is located in the blue wing of a strong Fe\ione\ 3819.62~\AA\ line, and the S/N in the blue wavelength range is lower. The above facts lead to uncertainty in the continuum placement and the derived abundance. NLTE calculations for Eu were performed with the Eu\ii\ model atom of \citet{2000AA...364..249M}. The NLTE abundance corrections for Eu\ii\ lines are available in the INASAN database\footnote{https://spectrum.inasan.ru/nLTE2/} \citep{2023MNRAS.524.3526M}. In the observed spectra of half of the sample stars, no detectable lines of Eu\ii\ were found, and we estimate upper limits on their Eu abundances. Our measurements for individual lines are given in Table~\ref{eu_fe_lbl}.

\begin{table}
\caption{NLTE and LTE abundances and EWs (m\AA) of Eu\ii\ lines in the sample stars.}
\label{eu_fe_lbl}
\centering   
   \begin{tabular}{llrrrr}    
      \hline
Name &  Eu\ii\  $\lambda$, \AA :  & &  3819 &    4129 &   4205 \\
\hline
\tiny{BPS CS 22881--036       } &\tiny{NLTE}  & &     --  &   --0.64 &   --0.62 \\
\tiny{BPS CS 22881--036       } & \tiny{LTE}  & &     --  &   --0.75 &   --0.73 \\
\tiny{BPS CS 22881--036       } & \tiny{EW}   & &     --  &      8.3 &      7.2 \\
\tiny{BPS BS 16080--175       } &\tiny{NLTE} & &     --  &   --0.27 &   --0.14 \\
\tiny{BPS BS 16080--175       } & \tiny{LTE}  & &     --  &   --0.39 &   --0.25 \\
\tiny{BPS BS 16080--175       } & \tiny{EW}   & &     --  &     19.5 &     22.7 \\
\tiny{HD 196944               } &\tiny{NLTE} & &     --  &   --1.28 &   --1.20 \\
\tiny{HD 196944               } & \tiny{LTE}  & &     --  &   --1.39 &   --1.29 \\
\tiny{HD 196944               } & \tiny{EW}   & &     --  &     28.4 &     31.3 \\
\tiny{BPS CS 31062--012       } &\tiny{NLTE} & &     --  &      --  &   --0.33 \\
\tiny{BPS CS 31062--012       } & \tiny{LTE}  & &     --  &      --  &   --0.44 \\
\tiny{BPS CS 31062--012       } & \tiny{EW}   & &     --  &      --  &     12.2 \\
\tiny{J09124370+0216236 } &\tiny{NLTE} & &   --0.64 &     --  &      --  \\
\tiny{J09124370+0216236 } & \tiny{LTE}  & &   --0.76 &     --  &      --  \\
\tiny{J09124370+0216236 } & \tiny{EW}   & &     16.5 &     --  &      --  \\
\tiny{BD --01 2582            } &\tiny{NLTE} & &   --0.99 &     --  &      --  \\
\tiny{BD --01 2582            } & \tiny{LTE}  & &   --1.09 &     --  &      --  \\
\tiny{BD --01 2582            } & \tiny{EW}   & &     64.4 &     --  &      --  \\
 \hline
\tiny{HE 1029-0546            } & & $<$ & \multicolumn{3}{c}{--2.11} \\ 
\tiny{HE 0450-4902            } & & $<$ & \multicolumn{3}{c}{--0.37} \\ 
\tiny{HE0024--2523            } & & $<$ & \multicolumn{3}{c}{--0.66} \\ 
\tiny{J11140709+1828320 }       & & $<$ & \multicolumn{3}{c}{--1.69} \\ 
\tiny{J11432342+2020582 }       & & $<$ & \multicolumn{3}{c}{--0.49} \\ 
 \hline
      \end{tabular}\\
\end{table}

\subsection{Abundance uncertainties}

We calculated the uncertainties in abundance ratios including uncertainties in stellar atmosphere parameters and observed spectra, as well as the dispersion of the individual line measurements around the mean $\sigma_{\rm st} = \sqrt{ \Sigma (\eps - \eps_i )^2 /(N - 1)}$, where N is the total number of lines. For spectra of different sample stars, their S/N vary from 32 to 266 (Table~\ref{sample}), and we estimated the error caused by continuum placement for each sample star. This value is negligible in spectra with S/N $>$ 200, while it reaches 0.13~dex in the spectrum of HE~0450-4902 with the lowest S/N of 32. For model atmosphere with \teff /log~g/[Fe/H]/\vt\ = 6140~K/4.38/$-1.9$/1.2\kms, we estimated the shifts in abundances from different lines caused by variations in \teff\ and \vt\ of 80~K and 0.1 \kms\ (Table~\ref{impact_lines}).

\begin{table}
\caption{Impact of uncertainties in stellar parameters on abundances from different lines}
 \label{impact_lines}
\centering   
   \begin{tabular}{lrrrrr}    
      \hline
 &  \multicolumn{3}{c}{Ba\ii} &  Eu\ii\ & Fe\ii  \\ 
$\Delta$ &   4934  & 5853 &  6141  & 4129  & 5018 \\  
   \hline 
\vt, $-0.1$ \kms & +0.03 & +0.05 & +0.03 & 0 & +0.04 \\
\teff, +80 K  & +0.08 & +0.06 & +0.08 & +0.05 & +0.01 \\
log~g, $-0.1$    & 0     & --0.02 & +0.01 & --0.04 & --0.03\\
 \hline
      \end{tabular}\\
\end{table}

In our Fe, Ba, and Eu abundance determination, we assume that the dispersion and  systematic errors caused by variations in stellar parameters and observatons are uncorrelated. This assumption is justified because the spectral lines of each of the studied elements belong to the same species, have nearly the same intensity, and originate from atomic levels with similar or identical \eexc. Therefore, changes in \teff, log~g, and \vt\  do not contribute to line-to-line scatter, allowing the total abundance ratio [X/H]  uncertainty, to be calculated as follows:\\
$\sigma^2_{\rm [X/H]} = \sigma_{\rm st}^2(X) + \sigma^2(T_{\rm eff}) + \sigma^2(\log~g) + \sigma^2(\xi_{\rm t}) + \sigma^2_{\rm obs}$

When computing the abundance ratios [X/Y], where X and Y are Fe, Ba and Eu, the uncertainties caused by changes in \teff\ , log~g, and \vt\ are reduced compared to those for the absolute abundances, since changes in parameters affect the Fe\ii , Ba\ii , and Eu\ii\ spectral lines in the same way. 
For abundance ratios X and Y, the uncertainty in their ratio is calculated as follows:\\
$\sigma^2_{\rm [X/Y]} = \sigma_{\rm st}^2(X) + \sigma_{\rm st}^2(Y) + \sigma^2(\Delta T_{\rm eff}) + \sigma^2(\Delta \log~g) + \sigma^2(\Delta \xi_{\rm t}) + \sigma^2_{\rm obs}$

The same is true for the uncertainty in the difference between the NLTE abundances from the subordinate and the resonance Ba\ii\ lines $\Delta$Ba(sub.-res.).
For illustration, we determined F$_{\rm odd}$ in HD~196944 using an additional set of stellar parameters \teff /log~g/[Fe/H]/\vt\ = 5539~K/2.44/$-2.50$ from \citet{2023A&A...679A.110G}. In our test calculations, we found F$_{\rm odd}$ = 0.64$_{-0.32}^{+0.36}$, which is consistent within the error bars with F$_{\rm odd}$ = 0.44$_{-0.22}^{+0.43}$, determined when using stellar parameters from this study. Both sets of stellar parameters yield similar [Ba/Fe] = 0.37 and 0.39 for parameters of \citet{2023A&A...679A.110G} and ours, respectively.

\subsection{Fractions of odd Ba isotopes}

\begin{figure}
\centering
\includegraphics[trim={0 5 0 0},width=\hsize]{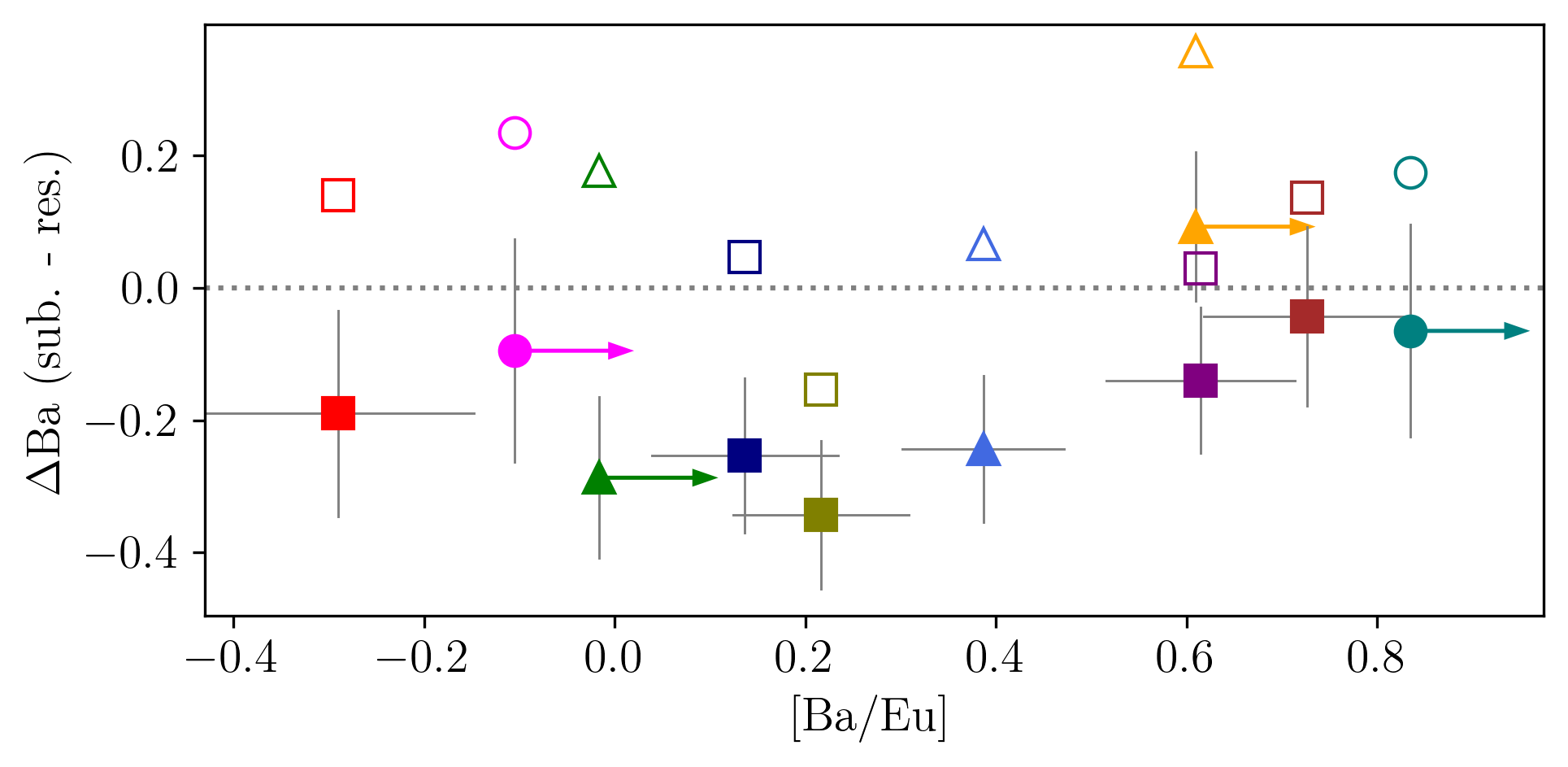}
\includegraphics[trim={0 5 0 0},width=\hsize]{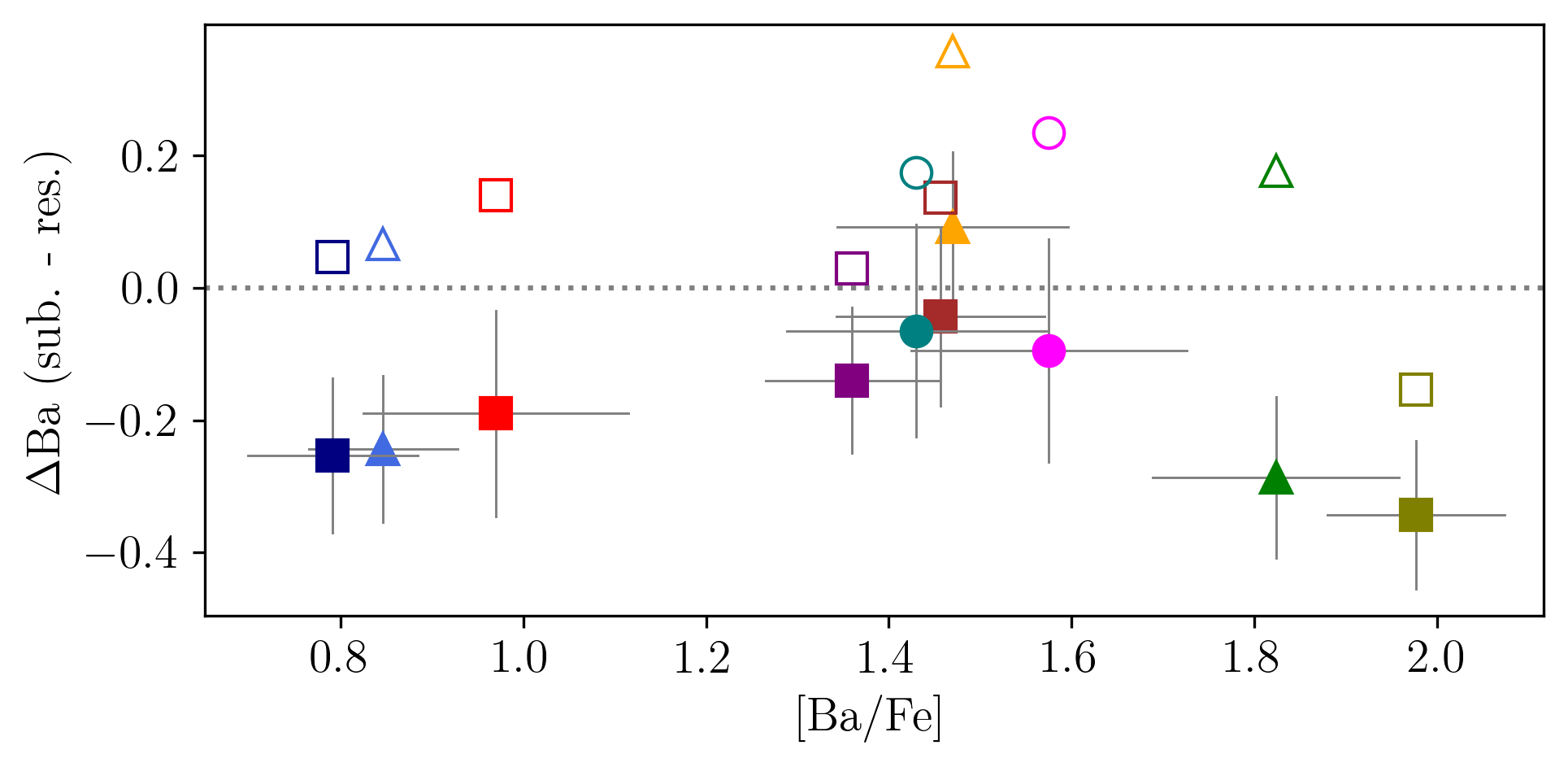}
\caption{Differences between NLTE abundances from the subordinate and the resonance Ba\ii\ lines computed with F$_{\rm odd}$ = 0.10 (s-process, filled symbols) and 0.75 (r-process, open symbols) in the sample stars known in the literature as CEMP-s (circles), CEMP-rs (squares), and with unclear classification (triangles).}
\label{ba_sub_res}
\end{figure}

Figure~\ref{ba_sub_res} shows $\Delta$Ba(sub.-res.) computed with F$_{\rm odd}$ = 0.10 (s-process) and 0.75 (r-process) in the sample stars.
Using the $\Delta$Ba(sub.-res.) computed for different F$_{\rm odd}$ from 0.1 to 1.0, we estimated F$_{\rm odd}$ that ensures equal NLTE abundances from the subordinate and resonance lines $\Delta$Ba(sub.-res.) = 0. The errors in F$_{\rm odd}$ were inferred  from the errors in $\Delta$Ba(sub.-res.), namely the maximum and the minimum F$_{\rm odd}$ ensure $\Delta$Ba(sub.-res.) = $\pm$ $\sigma_{\rm \Delta Ba(sub.-res.)}$.

Table~\ref{ratios_table} presents the derived abundance ratios and fractions of odd Ba isotopes together with their uncertainties. We took solar abundances from \citet{2021SSRv..217...44L}: $\eps$(Fe)$_{\rm \odot}$ = 7.45, $\eps$(Ba)$_{\rm \odot}$ = 2.17, and $\eps$(Eu)$_{\rm \odot}$ = 0.51.

\begin{table*}
\caption{NLTE abundance ratios and fractions of odd Ba isotopes in the sample stars}
\setlength{\tabcolsep}{0.90mm}
 \label{ratios_table}
\centering   
   \begin{tabular}{llrlrrrrlll}    
      \hline
Name & [Ba/H] &  \multicolumn{2}{c}{[Eu/H]} &  \multicolumn{2}{c}{[Ba/Eu]} & [Ba/Fe] & \multicolumn{2}{c}{[Eu/Fe]} & [Fe/H] & F$_{\rm odd}$ \\ 
   \hline 
\multicolumn{11}{c}{ F$_{\rm odd}$ is consistent with s- and inconsistent with r-process values} \\
2MASS J11140709+1828320   &  --1.59 (0.12) & $<$ &  --2.20 (0.22) & $>$ &   0.61 (0.23) &   1.47 (0.13) & $<$ &   0.86 (0.23) &  --3.06 (0.11)  & 0.05$_{-0.03}^{+0.07}$\\ 
HE 1029-0546              &  --1.79 (0.15) & $<$ &  --2.62 (0.24) & $>$ &   0.84 (0.24) &   1.43 (0.14) & $<$ &   0.59 (0.24) &  --3.22 (0.13)  & 0.19$_{-0.14}^{+0.50}$\\ 
HE 0450-4902              &  --0.99 (0.16) & $<$ &  --0.88 (0.25) & $>$ & --0.11 (0.25) &   1.58 (0.15) & $<$ &   1.68 (0.25) &  --2.56 (0.14)  & 0.19$_{-0.12}^{+0.33}$\\ 
\multicolumn{11}{c}{F$_{\rm odd}$ is consistent with r- and inconsistent with s-process values} \\
2MASS~J09124370+0216236  &  --1.44 (0.14) &     &  --1.15 (0.13) &     & --0.29 (0.14) &   0.97 (0.15) &     &   1.26 (0.14) &  --2.41 (0.12)  & 0.34$_{-0.21}^{+0.55}$\\ 
HD 196944                 &  --1.36 (0.09) &     &  --1.75 (0.08) &     &   0.39 (0.08) &   0.85 (0.08) &     &   0.46 (0.07) &  --2.21 (0.05)  & 0.44$_{-0.22}^{+0.43}$\\ 
BPS BS 16080-175          &  --0.10 (0.09) &     &  --0.72 (0.09) &     &   0.62 (0.10) &   1.36 (0.09) &     &   0.74 (0.10) &  --1.46 (0.07)  & 0.53$_{-0.38}^{+0.47}$\\
BD --01 2582              &  --1.36 (0.10) &     &  --1.50 (0.08) &     &   0.14 (0.10) &   0.79 (0.09) &     &   0.65 (0.08) &  --2.16 (0.06)  & 0.57$_{-0.31}^{+0.43}$\\
BPS CS 31062-012          &  --0.62 (0.09) &     &  --0.84 (0.08) &     &   0.22 (0.09) &   1.98 (0.10) &     &   1.76 (0.09) &  --2.60 (0.07)  &                       \\ 
\multicolumn{11}{c}{F$_{\rm odd}$ is consistent with both s- and r-process values} \\
BPS CS 22881-036          &  --0.41 (0.12) &     &  --1.14 (0.07) &     &   0.73 (0.11) &   1.46 (0.11) &     &   0.73 (0.06) &  --1.87 (0.06)  & 0.17$_{-0.14}^{+0.63}$\\ 
\multicolumn{11}{c}{F$_{\rm odd}$ is inconsistent with both s- and r-process values} \\
2MASS~J11432342+2020582 &  --1.02 (0.13) & $<$ &  --1.00 (0.23) & $>$ & --0.02 (0.23) &   1.82 (0.13) & $<$ &   1.84 (0.23) &  --2.84 (0.11)  & 0.36$_{-0.14}^{+0.23}$\\
 \hline
      \end{tabular}\\
Solar abundances are taken from \citet{2021SSRv..217...44L}. Total uncertainties are given in parenthesis. 
\end{table*}

\section{Discussion}\label{discussion}

\begin{figure}
\centering
\includegraphics[trim={0 5 0 0},width=\hsize]{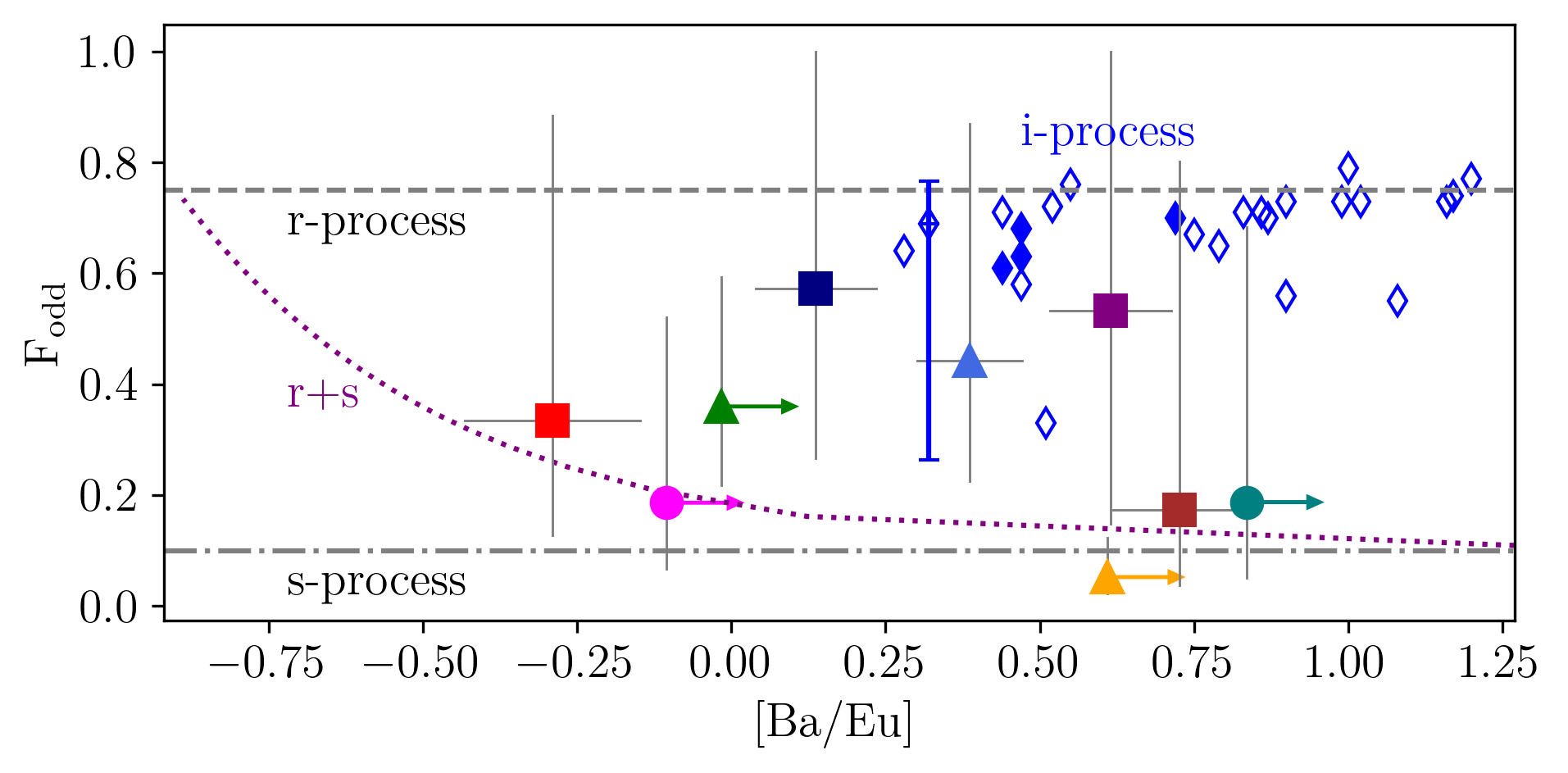}
\caption{F$_{\rm odd}$ as a function of [Ba/Eu] in the sample stars. Designations are the same as in Fig.~\ref{ba_sub_res}. For comparison, we show theoretical predictions for r-process (dashed line), s-process (dashdotted line), s- and r-process material mixture (dotted line), and i-process (diamonds). Filled diamonds correspond to i-process models with [Fe/H] and [Ba/Fe] that fall in the parameter range of our sample stars. Blue error bar indicates i-process nuclear uncertainties associated to 1~m$_{\rm \odot}$ model with [Fe/H] = $-2.5$.}
\label{baeu_fodd}
\end{figure}

\subsection{F$_{\rm odd}$ in different sample stars}

In three sample stars, we found consistent within the error bars NLTE abundances from the subordinate and the resonance lines when assuming the s-process  F$_{\rm odd}$, while the r-process value results in an abundance discrepancy larger than the error bars. Two of these stars, HE 1029-0546 and HE 0450-4902, are known in the literature as CEMP-s \citep{2015ApJ...807..173H}. Our Ba isotope analysis results in  F$_{\rm odd}$ = 0.19$_{-0.14}^{+0.50}$ and 0.19$_{-0.12}^{+0.33}$, respectively, which are close to solar value F$_{\rm odd}$ = 0.18, and thus confirms that barium in these stars originated mainly from the s-process. For the third star, 2MASS~J11140709+1828320, \citet{2013A&A...552A.107S} could not identify whether it is CEMP-s or CEMP-rs star, as it lacks Eu lines. Our analysis revealed that 2MASS~J11140709+1828320 is a CEMP-s star with [Ba/Eu] > 0.6 and the lowest F$_{\rm odd}$ = 0.05$_{-0.03}^{+0.07}$ among our sample stars. Hereafter, we refer to these three stars as CEMP-s stars.

For the other five sample stars, using the s-process  F$_{\rm odd}$ results in lower abundances from the subordinate lines compared to the resonance lines, and this cannot be explained by errors in the abundance determination method. In contrast to the s-process  F$_{\rm odd}$, using the r-process F$_{\rm odd}$ results in consistent within the error bars NLTE abundances from the subordinate and the resonance lines in these stars. An exception is  BPS~CS~31062-012, where we failed to estimate F$_{\rm odd}$, as the maximum F$_{\rm odd}$ = 1 provides $\Delta$Ba(sub.-res.) = $-0.13$~dex. This discrepancy is significantly smaller than that of $-0.34$~dex derived in BPS~CS~31062-012 when using the s-process F$_{\rm odd}$. Given the above information alongside with its [Ba/Eu] = 0.22, we consider this star to be a CEMP-rs star in agreement with its literature classification \citep{2015A&A...581A..22A}. In total, the four of these five stars are known as CEMP-rs stars, while the remaining star HD~196944 have an unclear classification. For HD~196944, we found [Ba/Eu] = 0.39 and F$_{\rm odd}$ = 0.44$_{-0.22}^{+0.43}$ and classified it as a CEMP-rs star. Hereafter, we refer to these five stars as CEMP-rs stars. The CEMP-rs sample stars exhibit [Ba/Eu] from $-0.3$ to 0.6, which is lower compared to the CEMP-s sample stars, which have lower limits on [Ba/Eu] $>$ 0.8, $>$ 0.6, and $>$ $-0.1$. 

Two sample stars remain undiscussed. One of them is BPS~CS~22881-036 known in the literature as CEMP-rs \citep{2015A&A...581A..22A}. For this star  with F$_{\rm odd}$ = 0.17$_{-0.14}^{+0.63}$ and [Ba/Eu] = 0.73 we found  that both s- and r-process F$_{\rm odd}$ yield consistent within the error bars   abundances from the subordinate and  the resonance lines. High [Ba/Eu] and close to solar  F$_{\rm odd}$ in BPS~CS~22881-036  align well with the values found in our CEMP-s sample stars.
The remaining star is SDSS~J114323.42+202058.0, where neither s- nor r- process F$_{\rm odd}$ are suitable for achieving abundance consistency. For this star we found F$_{\rm odd}$ = 0.36$_{-0.14}^{+0.23}$ and [Ba/Eu] $>$ 0. \citet{2013A&A...552A.107S} could not identify whether it is CEMP-s or CEMP-rs star, as it lacks Eu lines. Its  F$_{\rm odd}$ = 0.36 aligns well with those from 0.34 to 0.57 found in the CEMP-rs sample stars.

When plotting $\Delta$Ba(sub.-res.) as a function of [Ba/Fe] (Fig.~\ref{ba_sub_res}, bottom panel), we found that stars with a lower contribution of odd Ba isotopes to their Ba abundance have [Ba/Fe] of 1.5, while stars with a higher contribution of odd Ba isotopes have [Ba/Fe] of either $< 1.0$ or $> 1.8$. This finding aligns well with that of \citet{2010A&A...509A..93M}, see their Fig.~20, although the reason for this phenomenon remains unclear.  

\subsection{CEMP-rs stars: r+s processes or i-process}
In pure r-process material, F$_{\rm odd}$ = 0.75 and [Ba/Eu] = $-0.90$, while, in pure s-process material,  F$_{\rm odd}$ = 0.10 and [Ba/Eu] = 1.25 \citep{2020MNRAS.491.1832P}. Mixing them in different proportions results in a fixed relation between F$_{\rm odd}$ and [Ba/Eu] and higher [Ba/Eu] corresponds to a lower F$_{\rm odd}$. 

Figure~\ref{baeu_fodd} shows F$_{\rm odd}$ as a function of [Ba/Eu] in the sample stars, alongside with a curve that corresponds to a mixture of the s- and r-process material. We also indicate the r- and s-process predictions for F$_{\rm odd}$ taken from \citet{2020MNRAS.491.1832P} and the i-process predictions of \citet{2025EPJA...61...68C}. 
The i-process predictions correspond to individual AGB models with initial masses from 1 to 3 m$_{\rm \odot}$ and [Fe/H] from $-3$ to $-1$, for details see \citet{2025EPJA...61...68C}.

Among the four CEMP-rs sample stars, where we determined  F$_{\rm odd}$, three stars exhibit a discrepancy in  F$_{\rm odd}$ and [Ba/Eu] compared to those from the s- and r-process mixture: their high F$_{\rm odd}$ values of 0.44 -- 0.57 mismatch their high [Ba/Eu] of 0.14 -- 0.62. For comparison, equal r- and s-process material fractions yield  F$_{\rm odd}$ = 0.43 and [Ba/Eu] = $-0.60$. 
Therefore, mixing s- and r- process material cannot explain the observed F$_{\rm odd}$ and [Ba/Eu] in these CEMP-rs stars. 
We consider HD~196944, BPS~BS~16080$-$175, and BD$-$01~2582 as the strongest candidates thus far to encrypt i-process traces in their atmospheres.

The typical F$_{\rm odd}$  predicted for the i-process amounts to 0.6 -- 0.8 \citep{2025EPJA...61...68C}, although the nuclear uncertainties are large and, for example, an AGB model with [Fe/H] = $-2.5$ and a mass of 1~m$_{\rm \odot}$ yields F$_{\rm odd}$ = 0.69$_{-0.42}^{+0.07}$ \citep{2025EPJA...61...68C}. For six sample stars, we found their F$_{\rm odd}$ to be consistent within the error bars with the F$_{\rm odd}$ = 0.6 -- 0.8 predicted by the i-process. When accounting for the nuclear uncertainties, almost all our sample stars can be explained with the i-process. An exception is  2MASS~J11140709+1828320 star with  F$_{\rm odd}$ = 0.05$_{-0.03}^{+0.07}$ that clearly indicates its s-process origin.

\section{Conclusions}\label{conclusions}

We present a spectroscopic analysis of ten CEMP stars and determine their Ba and Eu NLTE abundances, as well as the fractions of odd Ba isotopes. For Ba isotope ratio analysis, we use a method based on abundance comparisons from the resonance and subordinate lines of Ba\ii. 
Our findings for the sample stars are summarised as follows:
\begin{itemize}
\item[$\circ$]
We found different F$_{\rm odd}$ in CEMP-s and CEMP-rs stars. 
\item[$\circ$]
CEMP-s stars exhibit F$_{\rm odd}$ = 0.05$_{-0.03}^{+0.07}$, 0.17$_{-0.14}^{+0.63}$, 0.19$_{-0.14}^{+0.50}$, and 0.19$_{-0.12}^{+0.33}$. Although the uncertainties are large, in three of four stars, the possibility of Ba isotopes origin in a pure r-process can be excluded. The obtained values agree, within the error bars, with the s-process F$_{\rm odd}$ = 0.10 and the solar F$_{\rm odd}$ = 0.18 predicted by \citet{2020MNRAS.491.1832P}.
\item[$\circ$]
CEMP-rs stars show higher F$_{\rm odd}$ compared to the CEMP-s stars with F$_{\rm odd}$ = 0.34$_{-0.21}^{+0.55}$, 0.36$_{-0.14}^{+0.23}$, 0.44$_{-0.22}^{+0.43}$, 0.53$_{-0.38}^{+0.47}$, and  0.57$_{-0.31}^{+0.43}$. Although the uncertainties are large, in four of five stars,  the possibility of a pure s-process origin for the Ba isotopes can be excluded. The obtained values agree, within the error bars, with the i-process F$_{\rm odd}$ = 0.6 to 0.8 predicted by \citet{2025EPJA...61...68C}.
\item[$\circ$]
In CEMP-rs stars with [Ba/Eu] > 0, their [Ba/Eu] and F$_{\rm odd}$ cannot be jointly explained by a mixture of material produced by the r- and s-processes.
\end{itemize}

The obtained results argue that the i-process is responsible for the chemical composition of the CEMP-rs sample stars. Further comprehensive abundance analysis of these stars is needed to obtain observational constraints on the i-process models.

\section{Data availability}
The full Tables are available at the CDS.

\begin{acknowledgements}
TS is grateful to Yu.~V.~Pakhomov for providing his code for echelle orders merging. 
Authors are grateful to Piercarlo Bonifacio and Elisabetta Caffau for providing their comments on the manuscript.
 
\end{acknowledgements}

\bibliography{aa57031-25}
\bibliographystyle{aa} 

\end{document}